\documentclass[pra,aps,11pt]{revtex4-2}
\usepackage{amsmath,amssymb,amsthm,mathtools,bm}
\usepackage{graphicx}
\usepackage{booktabs,tabularx,array}
\usepackage{microtype}
\usepackage{xcolor}
\usepackage{enumitem}
\usepackage[hidelinks]{hyperref}
\usepackage[T1]{fontenc}
\usepackage{lmodern}
\newtheorem{theorem}{Theorem}[section]

\newcommand{\Tr}{\operatorname{Tr}}
\newcommand{\diag}{\operatorname{diag}}
\newcommand{\Ree}{\operatorname{Re}}
\newcommand{\Imm}{\operatorname{Im}}
\newcommand{\ii}{\mathrm{i}}
\newcommand{\e}{\mathrm{e}}
\newcommand{\CS}{C_{\mathrm S}}
\newcommand{\hmd}{\mathrm{hom}}
\newcommand{\jy}[1]{\textcolor{black}{#1}}
\begin{document}
\title{Single-port homodyne detection achieves optimal multiphase sensing in an $n$-port ring cavity}
\author{Jiayu He}
\email{jiayu.he@helsinki.fi}
\affiliation{Department of Physics, University of Helsinki,
FI-00014 Helsinki, Finland}
\author{Aviral Kumar Pandey, Krishna Mohan Mishra, Devendra Kumar Mishra}
\email{dkmishra.phys@bhu.ac.in}
\affiliation{Department of Physics, Institute of Science, Banaras Hindu University, Varanasi 221005, India}
\author{Matteo G. A. Paris}
\email{matteo.paris@fisica.unimi.it}
\affiliation{Dipartimento di Fisica, Universit\`a di Milano,
I-20133 Milano, Italy}

\begin{abstract}
Estimating several phases simultaneously is more than estimating them one at a time: 
the optimal measurements for different parameters may not commute, and in a multiparameter 
problem the quantum Cram\'er-Rao bound is not automatically achievable. Ring cavities 
are a natural setting for this question, since a single coherent input can be made to 
interact repeatedly with every internal phase before it is read out. We study the simultaneous local estimation of the $n$ internal phases of a lossless $n$-port ring cavity probed by a fixed-energy multimode coherent state, optimizing over the full coherent-input space and over the operating point of the cavity. We identify a necessary and sufficient condition for the global optimum: the input energy should be distributed equally among the ports, and the internal phases and input phases satisfy a simple phase-matching condition for every port. The optimum is unique up to a common optical phase and minimizes both the \jy{symmetric logarithmic derivative (SLD)} cost  and the sloppiness of the model. At this point the quantum Fisher information matrix is full rank and its spectrum is obtained in closed form, the Uhlmann curvature vanishes, and no incompatibility penalty has to be paid. The precision retains the coherent-state scaling and the dependence on the number of phases, but the cavity feedback reduces the prefactor relative to $n$ independent single-pass experiments.
Finally, we show that this bound may be achieved in practice: the Fisher information of a 
balanced homodyne measurement at each output port, with local-oscillator phases locked 
to the prepared input phases is equal to the optimal quantum Fisher information.
{\color{black} Our results establish the ring cavity as a platform for optimal multiphase sensing and 
provide the ultimate benchmark against which nonclassical probes may be evaluated for further precision enhancement.
}
\end{abstract}
\maketitle
%%%
\section{Introduction}
Multiparameter quantum metrology represents a frontier in precision measurements, where the simultaneous estimation of multiple parameters, such as multiple phase-shifts in an interferometer, can offer advantages over estimating each parameter individually, especially when resources like total photon number are constrained\jy{~\cite{Humphreys2013,szczykulska2016,pezze2018,gessner2018,Albarelli2020,Ciampini2016,Polino2019,Hong2021}. Related developments include Gaussian probes, homodyne readout, and quantum-control approaches~\cite{Gagatsos2016,Gagatsos2022Erratum,
Rai2025Two,Rai2025Three,Yang_2022}}. The quantum Fisher information (QFI) establishes the ultimate precision limit for parameter estimation through the quantum Cramér–Rao bound. In the multiparameter setting, this role is taken by the quantum Fisher information matrix (QFIM), which captures the full precision capabilities of the system \cite{Helstrom1967,holevo2011}. A crucial complication, however, arises from the possible incompatibility of optimal measurements for different parameters, which introduces additional uncertainty. This incompatibility is characterized by the Uhlmann curvature and fundamentally affects the achievable joint estimation performance~\cite{Carollo2019,Albarelli2020,razavian2020quantumness}. 

Interferometric platforms offer natural settings for multiparameter estimation\jy{~\cite{Humphreys2013,pezze2018,gessner2018,
Ciampini2016,Polino2019,Hong2021},} with ring cavities being particularly interesting due to their ability to support multiple internal phases that can be estimated simultaneously\jy{~\cite{Belsley2022}.} Coherent states, while not exhibiting quantum-enhanced scaling, serve as excellent probes for interferometric applications due to their experimental accessibility and well-characterized properties~\cite{Caves1981,Giovannetti2011,Demkowicz2015}. Moreover, coherent states allow for closed-form analytical expressions for the QFIM and Uhlmann curvature, enabling a comprehensive theoretical analysis of the estimation problem. 

In this paper, we investigate the simultaneous local estimation of the $n$ internal phases of an $n$-port ring cavity using multimode coherent states. Under a fixed total incident energy, we optimize the probe over the complete coherent-input space. We obtain necessary and sufficient conditions for the global optimum and determine both the optimal coherent state and the corresponding cavity operating point. The optimal state is unique up to a common optical phase and simultaneously minimizes the SLD bound and the determinant-based sloppiness measure. We derive the complete spectrum of the optimal QFIM and show that it is full rank when an external phase reference is available. The Uhlmann curvature vanishes at the same point, so the optimal parameters are compatible. The precision retains the coherent-state scaling with the incident photon number and the number of estimated phases, but the ring cavity reduces the prefactor relative to separate single-pass estimation. Finally, we construct a port-resolved homodyne measurement whose classical Fisher information matrix equals the optimal QFIM. The predicted quantum precision bound is therefore attainable with a feasible measurement and does not require a collective measurement of the output modes.

The paper is organized as follows. Section~2 reviews the quantum multiparameter-estimation framework used in the analysis. Section~3 presents the $n$-port ring-cavity model, and introduces an equivalent uniform-phase representation together with the discrete Fourier input modes of the cyclic network. Section~4 states the optimal-probe theorem and explains the optimization method. Section~5 evaluates the optimal QFIM spectrum and the Uhlmann curvature, and then compares it with separate-estimation strategies. Section~6 compares the QFIM with the classical Fisher information of port-resolved homodyne detection. Section~7 specializes the general results to the three-port triangular cavity. Section~8 summarizes the main results. The detailed proofs are provided in the appendices.

\section{Quantum multiparameter estimation framework}
\label{sec:framework}
Let $\bm\lambda=(\lambda_1,\ldots,\lambda_p)^{\mathsf T}$ be a vector of real parameters encoded in a quantum state $\varrho_{\bm\lambda}$. The symmetric logarithmic derivative (SLD) $L_\mu$ is defined by\cite{Helstrom1967,holevo2011,Paris2009}
\begin{equation}
\partial_\mu\varrho=\frac12(L_\mu\varrho+\varrho L_\mu) ,
\end{equation}
where $\partial_\mu=\frac{\partial}{\partial\lambda_\mu}$, and the QFIM is
\begin{equation}
Q_{\mu\nu}=\frac12\Tr[\varrho(L_\mu L_\nu+L_\nu L_\mu)].
 \label{eq:qfim_framework}
\end{equation}
For $M$ independent repetitions and a locally unbiased estimator $\widehat{\bm\lambda}$, the SLD Cram\'er--Rao inequality reads
\begin{equation}
 \operatorname{Cov}(\widehat{\bm\lambda}) \succeq\frac1M Q^{-1}.
 \label{eq:sld_matrix_bound}
\end{equation}

A measurement described by a positive-operator-valued measure $\{\Pi_x\}$ produces probabilities $p(x|\bm\lambda)=\Tr(\varrho_{\bm\lambda}\Pi_x)$ and the classical Fisher information matrix~\cite{Fisher1922,Paris2009}
\begin{equation}
 F^{\mathrm C}_{\mu\nu} =\int dx\,p(x|\bm\lambda) \partial_\mu\ln p(x|\bm\lambda) \partial_\nu\ln p(x|\bm\lambda).
 \label{eq:classical_fim_framework}
\end{equation}
The QFIM bounds the information available from any measurement, \jy{but optimal measurements for different parameters need not coincide, so Eq.~\eqref{eq:sld_matrix_bound} is not automatically attainable~\cite{ragy2016,Pezze2017,Demkowicz2020,Hong2022,LuWang2021}. The distinction is also relevant to practical optical readout~\cite{Hong2022}.}

For a pure state $\varrho=\lvert\psi\rangle\langle\psi\rvert$, introduce the quantum geometric matrix~\cite{Provost1980}
\begin{equation}
J_{\mu\nu} =\langle\partial_\mu\psi\rvert(I-\lvert\psi\rangle\langle\psi\rvert)\lvert\partial_\nu\psi\rangle.
\label{eq:pure_geometry_framework}
\end{equation}
Its real and imaginary parts give
\begin{equation}
Q=4\Ree J, \qquad U=4\Imm J,
 \label{eq:geometry_framework}
\end{equation}
where $U$ is the Uhlmann curvature. Equivalently,
\begin{equation}
 U_{\mu\nu} =\frac{1}{2\ii}\Tr[\varrho(L_\mu L_\nu-L_\nu L_\mu)].
\end{equation}
$U=0$ is the weak-commutativity condition~\cite{ragy2016}, which removes an obstruction to simultaneous attainability of the SLD bound.

To compare QFIMs, a positive weight matrix $W$ defines the scalar SLD cost~\cite{ragy2016,Paris2009}
\begin{equation}
C_{\mathrm S}(W,Q)=\Tr(WQ^{-1}).
\end{equation}
We use the identity weight $W=I_p$ and write
\begin{equation}
\CS(Q)=\Tr(Q^{-1}).
\label{eq:CS_framework}
\end{equation}
This quantity is the sum of the local variance bounds in the chosen parameter coordinates. We also use~\cite{Yang_2022,he2025}
\begin{equation}
s(Q)=\frac1{\det Q},
 \label{eq:sloppiness_framework}
\end{equation}
which depends on the product of all information eigenvalues and therefore strongly penalizes a poorly resolved direction. If $Q$ is singular, at least one parameter combination cannot be resolved locally.

\section{Physical ring-cavity estimation model}
\label{sec:model}
\subsection{Beam splitters, propagation, and repeated circulation}
The cavity consists of $n\geq3$ identical lossless beam splitters connected along a directed ring. Indices are cyclic over $\{1,\ldots,n\}$. At beam splitter $j$, $a_j$ and $b_j$ are the annihilation operators of the external input and output modes, while $d_j$ and $c_{j\oplus1}$ are those of the incident and outgoing intracavity modes. Operator hats are omitted throughout for simplicity. We collect these operators into the vectors $\bm a,\bm b,\bm c,\bm d$. With power reflectivity $0<\rho<1$ and transmissivity $\tau=1-\rho$, we use~\cite{Paris_ring,Caves1}
\begin{align}
b_j&=\sqrt\rho\,a_j+\sqrt\tau\,d_j,\label{eq:local_bs_b}\\
c_{j\oplus1}&=\sqrt\tau\,a_j-\sqrt\rho\,d_j.\label{eq:local_bs_c}
\end{align}
\jy{The relative minus sign reflects the chosen beam-splitter phase
convention and ensures unitarity of the lossless transformation.} Propagation along the arm ending at port $j$ adds the unknown phase $\phi_j$,
\begin{equation}
d_j=\e^{\ii\phi_j}c_j.
\label{eq:propagation_matching}
\end{equation}

\begin{figure}[t]
\centering
\includegraphics[width=0.5\textwidth]{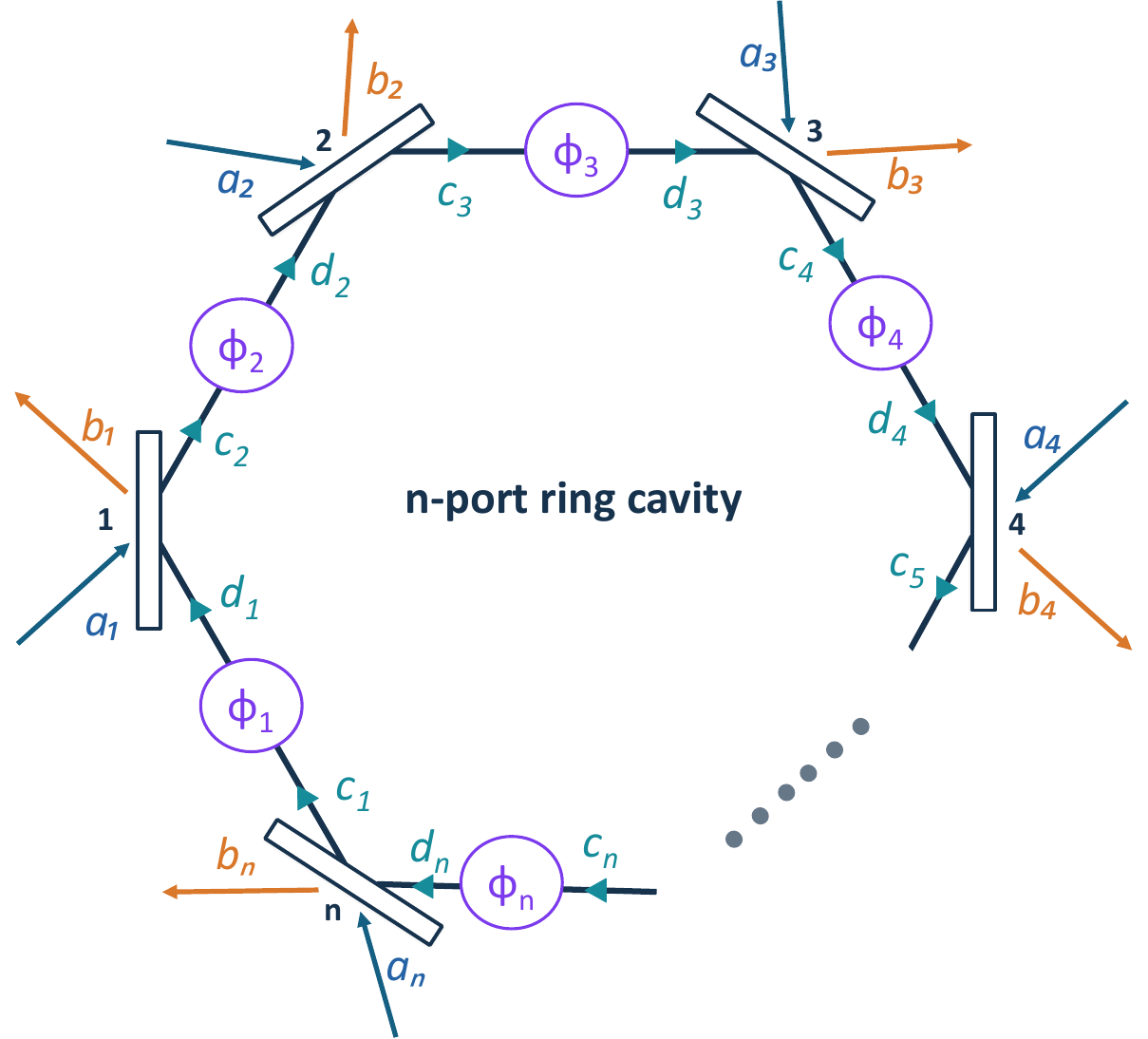}
\caption{Schematic $n$-port ring cavity. Each node is a beam splitter with an external input $a_j$, an output $b_j$, while $c_j$ and $d_j$ denote the internal modes. The arm ending at port $j$ contributes the phase $\phi_j$.}
\label{fig:ring_schematic}
\end{figure}

\jy{Fig~\ref{fig:ring_schematic} illustrates the propagation of light through the ring. At each port $j$, the external input field interferes with the circulating field arriving from the preceding arm at a beam splitter. One outgoing field leaves the ring through the external output port, while the other propagates along the next arm and acquires the phase $\phi_j$. It then reaches the next beam splitter, where it interferes with the external input at that port. Repeating this process around the ring closes the feedback loop, allowing the light to encounter the same phase elements on successive round trips. The $n$ propagation phases, one on each arm, form the unknown parameter vector $\bm\phi=(\phi_1,\ldots,\phi_n)^{\mathsf T}$, which we estimate jointly from the external output fields. To describe propagation around the ring in matrix form, we introduce the backward cyclic-shift matrix
$P$:}
\begin{equation}
P_{jk}=\delta_{k,j\ominus1},\qquad P^n=I_n,
\end{equation}
and define the single-step propagation matrix
\begin{equation}
T_{jk}=\e^{\ii\phi_j}\delta_{k,j\ominus1}.
\label{eq:T_physical_article}
\end{equation}
One complete circulation gives
\begin{equation}
T^n=\e^{\ii\Phi}I_n,\qquad  \Phi=\sum_{j=1}^{n}\phi_j.
\label{eq:round_trip_phase}
\end{equation}
Eliminating the intracavity modes yields
\begin{equation}
\bm b=S_n\bm a,\qquad S_n=(\sqrt\rho I_n+T)(I_n+\sqrt\rho T)^{-1}.
\label{eq:scattering_master_article}
\end{equation}
Losslessness implies $S_n^\dagger S_n=I_n$. The matrix
\begin{equation}
R=(I_n+\sqrt\rho T)^{-1}
\label{eq:R_definition_article}
\end{equation}
is the multiple-round-trip propagator: its finite expansion sums all directed paths within one circulation, while its scalar denominator \jy{accounts for} complete round trips. The derivation of Eq.~\eqref{eq:scattering_master_article} and the finite expansion of $R$ are given in Appendix~\ref{app:ring_algebra}.

\subsection{QFIM and Uhlmann curvature}
The input is a product coherent state
\begin{equation}
\lvert\bm\alpha\rangle=\bigotimes_{j=1}^{n}\lvert\alpha_j\rangle, \qquad \bm\alpha^\dagger\bm\alpha=N.
\label{eq:coherent_resource_article}
\end{equation}
Here $\alpha_j=\langle a_j\rangle$ is the coherent amplitude at input port $j$, and $N$ is the total mean input photon number. We denote the output amplitudes by $\beta_j=\langle b_j\rangle$. Linear optics preserves coherence~\cite{Caves1,Caves1981}, so the output state is
\begin{equation}
\lvert\psi(\bm\phi)\rangle=\bigotimes_{j=1}^{n}\lvert\beta_j(\bm\phi)\rangle, \qquad \bm\beta=S_n\bm\alpha.
\label{eq:coherent_output_article}
\end{equation}
The phases are estimated locally through the changes of the complex output amplitudes. Define
\begin{equation}
D_{j\mu}=\frac{\partial\beta_j}{\partial\phi_\mu},\qquad E_\mu=\bm e_\mu\bm e_\mu^{\mathsf T}.
\end{equation}
\jy{To quantify the response of the output fields to a change in $\phi_\mu$, we differentiate the output amplitudes while keeping the input fixed. Here $\partial_\mu T=\ii E_\mu T$, where $E_\mu$ projects onto the $\mu$th component. Since $R=(I+\sqrt{\rho}\,T)^{-1}$, its derivative is $\partial_\mu R =-\sqrt{\rho}\,R(\partial_\mu T)R,$ which accounts for the phase dependence accumulated over repeated
round trips. Combining these derivatives gives}
\begin{equation}
\partial_\mu S_n=\ii\tau R E_\mu T R.
\label{eq:S_derivative_article}
\end{equation}
Consequently,
\begin{equation}
D=\ii\tau R\diag(TR\bm\alpha).
\label{eq:D_factorization_article}
\end{equation}
The column form makes the optical mechanism explicit:
\begin{equation}
D_{\cdot\mu}=\ii\tau (TR\bm\alpha)_\mu R\bm e_\mu.
\label{eq:D_column_physical}
\end{equation}
A variation of $\phi_\mu$ first creates a local source proportional to the field $(TR\bm\alpha)_\mu$ arriving at that phase element. The propagator $R$ then distributes this response among all output ports.

For a multimode coherent state, the quantum geometric matrix is the Gram matrix of these response columns,
\begin{equation}
J=D^\dagger D.
\label{eq:complex_gram_article}
\end{equation}
Hence the QFIM and Uhlmann curvature are
\begin{equation}
Q=4\Ree J, \qquad U=4\Imm J.
\label{eq:Q_U_article}
\end{equation}
Appendix~\ref{app:coherent_geometry} derives Eq.~\eqref{eq:complex_gram_article} directly from coherent-state derivatives.

\subsection{Equivalent cyclic representation and winding modes}
At a fixed operating point $\bm\phi_0$, an arbitrary distribution of internal phases can be converted into a uniform cyclic form by a diagonal rephasing. There exists
\begin{equation}
G=\diag(\e^{\ii\gamma_1},\ldots,\e^{\ii\gamma_n})
\end{equation}
such that
\begin{equation}
G^\dagger T_0G=\e^{\ii\Phi_0/n}P,\qquad \Phi_0=\sum_j\phi_{0,j}.
\label{eq:gauge_reduction_article}
\end{equation}
This is a fixed change of mode coordinates in the local estimation problem, not a restriction on the cavity or probe. Let $\bm\alpha_{\mathrm{phys}}$ denote the physical coherent-amplitude vector. In the transformed mode coordinates,
\begin{equation}
\widetilde{\bm\alpha}=G^\dagger\bm\alpha_{\mathrm{phys}},\qquad \widetilde D=G^\dagger D.
\end{equation}
Since $G$ is unitary,
\begin{equation}
\widetilde D^\dagger\widetilde D=D^\dagger D,\qquad \widetilde{\bm\alpha}^\dagger\widetilde{\bm\alpha} = \bm\alpha_{\mathrm{phys}}^\dagger\bm\alpha_{\mathrm{phys}}=N.
\end{equation}
Thus every physical coherent input corresponds bijectively to a transformed input with the same total energy. We define its normalized input profile by
\begin{equation}
\bm z:= \frac{\widetilde{\bm\alpha}}{\sqrt N}, \qquad \bm z^\dagger\bm z=1.
\label{eq:normalized_transformed_input}
\end{equation}
The vector $\bm z$ specifies the relative amplitudes and phases of a particular coherent input in the transformed mode coordinates. The Fourier vectors introduced below will provide an orthonormal basis in which an arbitrary $\bm z$ can be expanded. Because $G$ is fixed at the operating point and unitary, $\widetilde J=\widetilde D^\dagger\widetilde D=J$; the transformation also preserves the input energy, so $Q$, $U$, $\CS$, and $s$ are unchanged.

From this point onward, we work entirely in the transformed coordinates and omit both the tildes and the operating-point subscript. The model is written as
\begin{equation}
T=\e^{\ii\Phi/n}P,\qquad \bm\alpha=\sqrt N\,\bm z,\qquad \bm z^\dagger\bm z=1.
\label{eq:uniform_coordinates_article}
\end{equation}
Appendix~\ref{app:existenceG} proves the existence of $G$ and shows explicitly why it remains fixed when the local derivatives are taken. The cyclic structure of the transformed propagation matrix suggests using the discrete Fourier basis of the mode space. Define $\omega=\e^{2\pi\ii/n}$ and the Fourier vectors
\begin{equation}
(\bm z_m)_j= \frac{1}{\sqrt n} \omega^{-m(j-1)}, \qquad j=1,\ldots,n,
 \label{eq:Fourier_input_modes}
\end{equation}
for $m=0,\ldots,n-1$. They form an orthonormal basis of the transformed input-mode space satisfying $\bm z_m^\dagger\bm z_{m'}=\delta_{mm'}$ and $\sum_{m=0}^{n-1}\bm z_m\bm z_m^\dagger=I_n$. They reproduce themselves after one cyclic step,
\begin{equation}
P\bm z_m=\omega^m\bm z_m, \qquad T\bm z_m=\e^{\ii\Upsilon_m}\bm z_m,
\label{eq:effective_increment_article}
\end{equation}
where $$ \Upsilon_m=\frac{\Phi+2\pi m}{n}.$$ In the uniform cyclic coordinates, $m$ labels a Fourier eigenvector $\bm z_m$ of $P$, while $j$ labels an input port. For a fixed $m$, the component $(\bm z_m)_j$ gives the normalized complex input amplitude at port $j$. Different values of $m$ give different phase differences between neighboring ports, while the input intensity remains equal at every port. We call these spatial phase patterns winding modes.

The vectors $\{\bm z_m\}$ form a complete orthonormal basis, so an arbitrary normalized input has the unique expansion
\begin{equation}
\bm z=\sum_{m=0}^{n-1}c_m\bm z_m,\qquad \sum_{m=0}^{n-1}|c_m|^2=1.
\label{eq:arbitrary_winding_expansion}
\end{equation}
\jy{The index $m$ labels different input phase configurations: each mode $\bm z_m$ has equal intensity $1/n$ at all ports, but a different phase difference between adjacent inputs. The expansion above allows
us to express an arbitrary input in terms of these modes.}

\section{Optimal coherent-state probe}
\label{sec:optimality}

The optimization of $\CS$ and $s$ over the internal phase configuration and the complete fixed-energy coherent-input space admits the same global optimum, summarized in the following theorem.

\begin{theorem}[Global optimum over fixed-energy coherent inputs]
\label{thm:global_article}
Consider a lossless $n$-port ring cavity with $n\geq3$ and $0<\rho<1$. For simultaneous local estimation of all $n$ internal phases with an external phase reference, the costs $\CS=\Tr(Q^{-1})$ and $s=(\det Q)^{-1}$ attain their global minimum over all coherent inputs satisfying $\bm\alpha^\dagger\bm\alpha=N>0$ and all internal phase operating points when the effective phase increment is
\begin{equation}
\Upsilon_*=\pi\pmod{2\pi}.
\label{eq:Upsilon_opt_article}
\end{equation}
Writing the physical input amplitudes as $\alpha_j=|\alpha_j|\e^{\ii\theta_j}$, the minimizing configurations are characterized by
\begin{equation}
|\alpha_j|^2=\frac{N}{n},\qquad \phi_j+\theta_{j\ominus1}-\theta_j=\pi\pmod{2\pi}
\label{eq:physical_optimum_article}
\end{equation}
for every port $j$. For any fixed internal phase distribution satisfying $\Phi_*=n\pi\pmod{2\pi}$, Eq.~\eqref{eq:physical_optimum_article} uniquely determines the relative phases of the minimizing input, leaving only an arbitrary common input phase.
\end{theorem}

% At the optimum, direct substitution into the scattering relation yields $\bm\beta_{\mathrm{opt}}=-\bm\alpha_{\mathrm{opt}}$. The optimal output therefore has the same spatial mode profile and energy distribution as the input, differing only by a common phase $\pi$.

We optimize the estimation problem under the constraint of a fixed total photon number $N$. The primary cost is the SLD bound $\CS(Q)=\operatorname{Tr}(Q^{-1})$, which bounds the sum of the estimation variances. We also consider the sloppiness $s(Q)=(\det Q)^{-1}$, which depends on the product of the information carried by all parameter directions and becomes large when the QFIM contains a poorly estimated direction. Although these two costs quantify precision differently, they share the properties needed below: they are invariant under cyclic relabeling, convex on the set of positive-definite QFIMs, and monotonically decreasing under the Loewner order. Consequently, the same optimization argument applies to both costs, and more generally to any function $L(Q)$ with these properties.

Let $\boldsymbol z$ be an arbitrary normalized coherent input in the uniform cyclic coordinates, and let $P^r\boldsymbol z$ denote its $r$-step cyclic shift. The QFIM of each cyclically shifted input is related to the original QFIM by $Q(P^r\boldsymbol z)=P^rQ(\boldsymbol z)P^{-r}$. Therefore, all QFIMs in this cyclic family have the same eigenvalues and the same values of $\CS$ and $s$. We introduce the comparison matrix
$$ \overline Q(\boldsymbol z)=\frac1n\sum_{r=0}^{n-1}Q(P^r\boldsymbol z). $$
This matrix is not the QFIM of the original input; it is the average of the QFIMs obtained from all cyclic shifts of that input. Convexity and cyclic invariance give $$L(\overline Q)\leq \frac{1}{n}\sum _{r=0}^{n-1}L(P^rQ(\boldsymbol z)P^{-r})=L(Q(\boldsymbol z)).$$ Expanding the input in the discrete Fourier basis, $\boldsymbol z=\sum_m c_m\boldsymbol z_m$, shows why this average is useful: it cancels the cross terms between distinct Fourier components and gives $\overline Q=\sum_m|c_m|^2Q_m$, where $Q_m=Q(\boldsymbol z_m)$. \jy{All $Q_m$ have the same parameter-space Fourier eigenvectors, with eigenvalues $\lambda_k^{(m)}$. Here $m$ labels the input mode, while $k$ labels a Fourier mode of small variations in the internal phases. Thus $\lambda_k^{(m)}$ quantifies the information about phase-variation mode $k$ for input $\bm z_m$. The eigenvalues satisfy $\lambda_k^*\geq\lambda_k^{(m)}$ for every $m$ and $k$. Thus, among Fourier inputs, the optimal mode maximizes the information about every phase-variation mode simultaneously.} Hence $Q_*\succeq Q_m$ for every input Fourier mode and therefore $Q_*\succeq\overline Q$. Monotonicity under the Loewner order then yields the central comparison
$$L(Q_*)\leq L(\overline Q)\leq L(Q(\boldsymbol z)).$$
 
The equality conditions identify both the optimal cavity operating point and the optimal input. The eigenvalue inequality in the common-phase direction is strict whenever $\Upsilon_m\neq\pi$. Equality in the above chain requires the Fourier expansion of the input to contain only the single mode $\boldsymbol z_{m_*}$ satisfying $\Upsilon_{m_*}=\pi$, so that $\boldsymbol z=e^{i\chi}\boldsymbol z_{m_*}$ for an arbitrary common phase $\chi$. The corresponding total internal phase is $\Phi_*=n\pi\pmod{2\pi}$. Returning to the original physical coordinates gives Eq.~(\ref{eq:physical_optimum_article}). Thus the optimal probe has equal incident energy at all ports, while its input phases depend on the internal phase distribution. Conversely, any input satisfying these conditions is mapped to the optimal Fourier mode. 

The theorem establishes a necessary and sufficient condition for global optimality over the complete set of fixed-energy coherent inputs and internal phase configurations. The stated energy and phase conditions are sufficient to attain the minimum, while the strict inequality in the common-phase direction shows that no other coherent input can attain the same value. The same configuration minimizes both $\CS=\operatorname{Tr}(Q^{-1})$ and $s=(\det Q)^{-1}$, showing that the optimizer does not depend on which of these two precision criteria is used. Moreover, all modal QFIMs share the same parameter-space Fourier eigenvectors, and their eigenvalues satisfy $\lambda_k^*\geq\lambda_k^{(m)}$ for every input mode $m$ and every parameter direction $k$. Thus, within the modal comparison, the optimal configuration improves no direction at the expense of another. The complete proof is given in the Appendix~\ref{app:Proof_thm}.

\section{Optimal precision and incompatibility}
\label{sec:consequences}
\subsection{QFIM spectrum and scalar costs}
At $\Upsilon_*=\pi$, the eigenvalue of Q becomes
\begin{equation}
\lambda_k^*=\frac{4N(1+\sqrt\rho)^2}{n[1+\rho-2\sqrt\rho\cos(2\pi k/n)]}.
\label{eq:lambda_opt_article}
\end{equation}
Every eigenvalue is positive for $0<\rho<1$, so all $n$ local phase directions are identifiable. The full-rank estimation of all $n$ internal
phases is understood relative to a stable external optical phase reference shared by the coherent-state preparation and homodyne detection. Without such a reference, the identifiability of the common-phase direction
requires separate consideration. The spectrum is generally anisotropic: the uniform phase variation $k=0$, in which all internal phases change by the same amount, receives the strongest recirculation response, while the other spatial frequencies have smaller enhancement.

Using $\sum_k\cos(2\pi k/n)=0$ gives
\begin{equation}
\CS^*=\frac{n^2(1+\rho)}{4N(1+\sqrt\rho)^2}.
\label{eq:CS_opt_article}
\end{equation}
\jy{At fixed $\rho$, the optimal cost scales as $n^2/N$.
The cavity reflectivity affects only the prefactor without
changing the scaling with either the photon number $N$ or
the number of phases $n$, consistent with the constant-factor simultaneous-estimation advantage discussed in Ref.~\cite{Gorecki2022}.} The roots-of-unity product $\prod_{k=0}^{n-1}[1+\rho-2\sqrt\rho\cos(2\pi k/n)]=(1-\rho^{n/2})^2$ yields
%\begin{equation}
 %\det Q_*=\frac{[4N(1+\sqrt\rho)^2/n]^n}{(1-\rho^{n/2})^2},
%\end{equation}
%and
\begin{equation}
s^*=\left[\frac{n}{4N(1+\sqrt\rho)^2}\right]^n(1-\rho^{n/2})^2.
\label{eq:s_opt_article}
\end{equation}
The ring retains the coherent-state $1/N$ precision scaling; repeated circulation modifies its prefactor.

\subsection{Vanishing incompatibility}
Before taking the real part, the  quantum geometric matrix eigenvalues are
\begin{equation}
\Gamma_k^{(m)}=\frac{N(1-\rho)^2}{nf_mf_{m+k}}.
\label{eq:Gamma_article}
\end{equation}
The Uhlmann curvature eigenvalues are
\begin{equation}
u_k^{(m)}=2\ii[\Gamma_{-k}^{(m)}-\Gamma_k^{(m)}].
\label{eq:curvature_spectrum_article}
\end{equation}
At $\Upsilon=\pi$, $\Gamma_k^*=\Gamma_{-k}^*$ and
\begin{equation}
U_*=0.
\label{eq:U_zero_article}
\end{equation}
The precision-optimal configuration is therefore weakly compatible. It follows from cancellation of the oriented imaginary response between conjugate Fourier directions.

\subsection{Comparison with separate estimation}
\begin{figure}[t]
    \centering
    \includegraphics[width=0.7\columnwidth]
    {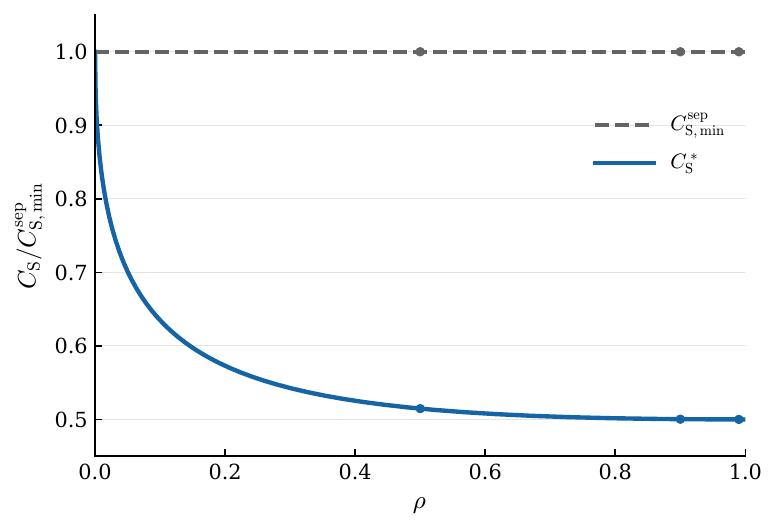}
    \caption{Optimal ring cost $C_{\mathrm S}^*$ relative to the minimum separate-estimation cost $C_{\mathrm S,\min}^{\mathrm{sep}}=n^2/(4N)$, as a function of the coupler power reflectivity $\rho$. The solid curve shows $C_{\mathrm S}^*/C_{\mathrm S,\min}^{\mathrm{sep}}=(1+\rho)/(1+\sqrt{\rho})^2$, while the dashed line denotes the separate-estimation benchmark. Cavity feedback reduces the cost by a constant factor, approaching $1/2$ as $\rho\rightarrow1$, without changing the $n^2/N$ scaling.}
    \label{fig:precision_comparison}
\end{figure}

As a fixed-incident-energy benchmark, suppose the phases are estimated in $n$ independent single-pass coherent experiments with photon numbers $N_j$ satisfying $\sum_jN_j=N$. The individual QFIs are $4N_j$, so
\begin{equation}
\CS^{\mathrm{sep}}=\sum_{j=1}^{n}\frac1{4N_j}\geq\frac{n^2}{4N}\jy{=:C_{\mathrm S,\min}^{\mathrm{sep}}},
\label{eq:separate_cost_article}
\end{equation}
with equality at $N_j=N/n$. Hence
\begin{equation}
\frac{\CS^*}{C_{\mathrm S,\min}^{\mathrm{sep}}}=\frac{1+\rho}{(1+\sqrt\rho)^2}<1.
\label{eq:separate_ratio_article}
\end{equation}
The advantage is a prefactor improvement under fixed incident energy, not a change of shot-noise scaling. Because the cavity generates repeated photon--phase interactions, a benchmark based on intracavity energy, residence time, or total interactions would require a different resource normalization. \jy{As shown in Fig.~\ref{fig:precision_comparison}, the improvement
increases monotonically with the coupler reflectivity and approaches a factor of two in the limit $\rho\rightarrow1$.}

\section{Homodyne attainment}
\label{sec:homodyne_attainment}

The QFIM provides a measurement-independent bound on estimation precision, but its simultaneous attainment is not guaranteed in a multiparameter problem. We now show that, at the optimal probe and operating point, one balanced homodyne measurement at each physical output port yields a classical Fisher information matrix equal to $Q_*$. The scheme requires no optical mixing between distinct output ports. Throughout this section, we assume ideal photodetection and the strong-local-oscillator limit.

\begin{figure}[t]
 \centering
\includegraphics[width=0.6\textwidth]{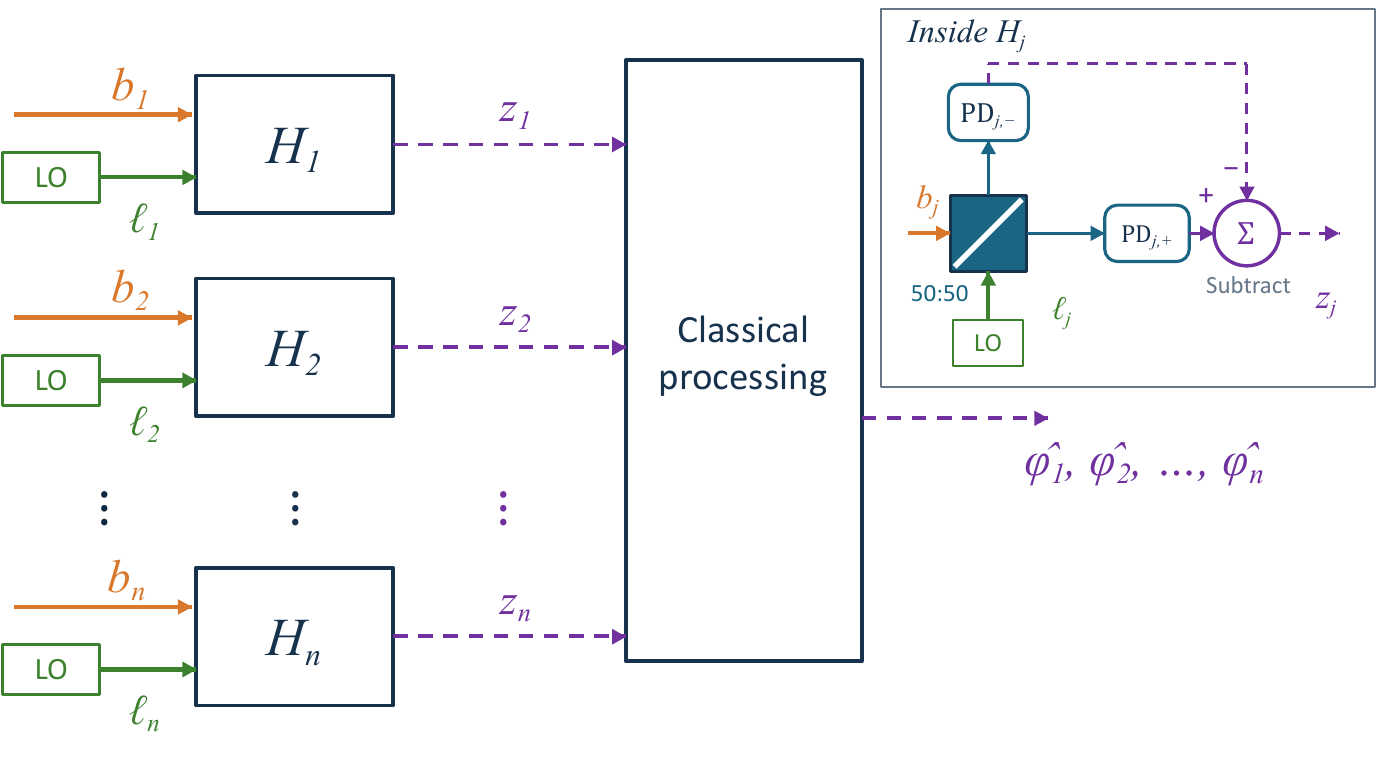}
 \caption{Direct homodyne readout of the $n$ cavity outputs. Each signal mode $b_j$ enters a balanced homodyne module $H_j$ together with a strong, phase-referenced local oscillator $\ell_j$, producing a quadrature record $z_j$. The inset shows the optical and electronic components of $H_j$: a $50{:}50$ beam splitter, two photodetectors $\mathrm{PD}_{j,\pm}$, and subtraction of their records followed by normalization. Solid lines denote optical paths and dashed lines denote electrical signals. Joint classical processing of the records yields estimates of the internal phases. At the optimal operating point, the local-oscillator phases satisfy Eq.~\eqref{eq:LO_physical_article}, with $\theta_j$ the corresponding prepared input phase.}
 \label{fig:homodyne_readout}
\end{figure}

The measurement scheme is illustrated in Fig.~\ref{fig:homodyne_readout}. At port $j$, the signal mode $b_j$ interferes with a coherent local oscillator $\ell_j$ on a $50{:}50$ beam splitter. The local-oscillator amplitude is $\langle\ell_j\rangle=B_j\e^{\ii\zeta_j}$, with $B_j>0$ and a calibrated phase $\zeta_j$. The two photodetection records are subtracted and normalized. In the strong-local-oscillator limit, the resulting value $z_j$ is an outcome of the quadrature
\begin{equation}
\begin{aligned}
 X_j(\zeta_j)=\frac{\e^{-\ii\zeta_j}b_j+\e^{\ii\zeta_j}b_j^\dagger}{\sqrt2},\quad z_j=\frac{n_{j,+}-n_{j,-}}{\sqrt2 B_j},
\end{aligned}
\label{eq:homodyne_quadrature_readout_article}
\end{equation}
where $n_{j,\pm}$ are the photon-number measurement outcomes at the two beam-splitter outputs $u_{j,\pm}=(b_j\pm\ell_j)/\sqrt2$. The corresponding photon-number observables are $$u_{j,\pm}^\dagger u_{j,\pm}=\frac12\left[b_j^\dagger b_j+\ell_j^\dagger\ell_j \pm\left(b_j^\dagger\ell_j+\ell_j^\dagger b_j\right)\right].$$ For the optimal physical input $\alpha_j^{\mathrm{opt}}=\sqrt{N/n}\,\e^{\ii\theta_j}$, the required local-oscillator phase is
\begin{equation}
\zeta_{j,\mathrm{phys}}^{\mathrm{opt}}=\theta_j+\frac{\pi}{2}\pmod\pi.
\label{eq:LO_physical_article}
\end{equation}
Thus each local oscillator is shifted by $\pi/2$ relative to the phase of the corresponding prepared input. All local oscillators share a phase reference with the probe preparation.

Since the output is a product of coherent states, the quadrature outcomes are independent Gaussian variables. Their means and variances are
\begin{equation}
 m_j=\sqrt2\Ree\!\left(\e^{-\ii\zeta_j}\beta_j\right), \qquad \operatorname{Var}(z_j)=\frac12,
\label{eq:homodyne_statistics_article}
\end{equation}
and the joint probability density is
\begin{equation}
p(\mathbf z\mid\phi_1,\ldots,\phi_n)=\prod_{j=1}^{n}\frac{\e^{-(z_j-m_j)^2}}{\sqrt\pi}.
\label{eq:homodyne_likelihood_article}
\end{equation}
At fixed probe and local-oscillator settings, the parameter dependence enters only through the means. Using $D_{j\mu}=\partial_\mu\beta_j$, the classical Fisher information matrix per repetition is
\begin{equation}
\begin{aligned}
 F_{\mu\nu}^{\hmd}=2\sum_{j=1}^{n}(\partial_\mu m_j)(\partial_\nu m_j)=4\sum_{j=1}^{n}\Ree\!\left(\e^{-\ii\zeta_j}D_{j\mu}\right)\Ree\!\left(\e^{-\ii\zeta_j}D_{j\nu}\right).
\end{aligned}
\label{eq:homodyne_FIM_article}
\end{equation}

The reason this readout suffices at the optimum is the phase structure of the response matrix. Appendix~\ref{app:homodyne_response} derives the optimal response and its expression in the physical port coordinates. In these coordinates, each response row has a common phase modulo $\pi$, so that
\begin{equation}
D_{j\mu}^{\mathrm{opt}}=\e^{\ii\zeta_{j,\mathrm{phys}}^{\mathrm{opt}}}r_{j\mu}, \qquad r_{j\mu}\in\mathbb R.
\label{eq:D_opt_rows_article}
\end{equation}
Consequently, all first-order changes of the coherent amplitude at a given port lie along the same quadrature direction. Measuring that quadrature retains the information carried by every parameter derivative at that port, even though an individual record generally depends on several phases.

To establish attainment, define $\Theta_{\boldsymbol\zeta}=\diag(\e^{-\ii\zeta_1},\ldots,\e^{-\ii\zeta_n})$. Combining Eq.~\eqref{eq:homodyne_FIM_article} with $Q=4\Ree(D^\dagger D)$ and the unitarity of $\Theta_{\boldsymbol\zeta}$ gives
\begin{equation}
\begin{aligned}
Q-F^{\hmd}=4\bigl[\Imm(\Theta_{\boldsymbol\zeta}D)\bigr]^{\mathsf T} \Imm(\Theta_{\boldsymbol\zeta}D)\succeq0.
\end{aligned}
\label{eq:homodyne_deficit_article}
\end{equation}
The right-hand side quantifies the information missed by the selected quadratures. For this readout, equality holds if and only if every element of $\Theta_{\boldsymbol\zeta}D$ is real. Eqs.~\eqref{eq:LO_physical_article} and~\eqref{eq:D_opt_rows_article} ensure this condition at the optimum, yielding
\begin{equation}
F^{\hmd}=Q_*.
\label{eq:homodyne_attainment_article}
\end{equation}
This attainment is local. The input phases and local-oscillator phases are calibrated at the chosen operating point, and all parameter derivatives are taken with the prepared probe and measurement settings held fixed. The transformation $G$ used in the derivation is a change of coordinates; Eq.~\eqref{eq:LO_physical_article} states the prescription directly in physical phases, so no implementation of $G$ is required. Under the usual regularity conditions for local estimation, $M$ independent repetitions and an asymptotically efficient classical estimator then give a covariance matrix approaching $Q_*^{-1}/M$.

\section{Triangular cavity as an example}
\label{sec:triangle}
The triangular cavity is the smallest ring with three independently encoded phases. It provides a direct physical realization of the general theorem without requiring a separate optimization proof.

\subsection{Optimal input and precision}
For $n=3$, the optimal total phase and winding branch are
\begin{equation}
\Phi_*=\pi, \qquad m_*=1, \qquad \Upsilon_*=\pi.
\end{equation}
The physical conditions are
\begin{equation}
|\alpha_1|^2=|\alpha_2|^2=|\alpha_3|^2=\frac N3,\qquad \phi_j+\theta_{j\ominus1}-\theta_j=\pi.
\label{eq:triangle_physical_conditions}
\end{equation}
Fixing the common input phase by $\theta_1=0$, one representative is
\begin{equation}
\theta_2=\phi_2-\pi,\qquad \theta_3=\pi-\phi_1,\qquad \phi_1+\phi_2+\phi_3=\pi.
\label{eq:triangle_input_phases}
\end{equation}
For the symmetric internal-phase representative $\phi_1=\phi_2=\phi_3=\pi/3$, the optimal input is $(\alpha_1,\alpha_2,\alpha_3)=\sqrt{\frac N3}(1,\e^{-2\pi\ii/3},\e^{2\pi\ii/3})$. Equal internal phases are not required; the input phases compensate any optimal internal distribution according to Eq.~\eqref{eq:triangle_physical_conditions}.

For three estimated phases, the optimal QFIM has one eigenvalue associated with a common variation of all internal phases and two degenerate eigenvalues associated with relative-phase variations. The common variation changes all three phases by the same amount, whereas the relative variations satisfy $\delta\phi_1+\delta\phi_2+\delta\phi_3=0$. Eq.~\eqref{eq:lambda_opt_article} gives
\begin{equation}
\lambda_{\mathrm{com}}^*=\frac{4N(1+\sqrt\rho)^2}{3(1-\sqrt\rho)^2},
\label{eq:triangle_common_eigenvalue}
\end{equation}
and
\begin{equation}
\lambda_{\mathrm{rel}}^*=\frac{4N(1+\sqrt\rho)^2}{3(1+\rho+\sqrt\rho)},
\label{eq:triangle_relative_eigenvalue}
\end{equation}
where $\lambda_{\mathrm{rel}}^*$ is doubly degenerate. For $0<\rho<1$, $\lambda_{\mathrm{com}}^*>\lambda_{\mathrm{rel}}^*>0$, so the cavity is more sensitive to a common change of the three phases, while the two independent relative-phase variations remain locally identifiable. The vanishing Uhlmann curvature establishes compatibility, whereas the unequal QFIM eigenvalues show that the phase sensitivities are not isotropic. The scalar results reduce to
\begin{equation}
\CS^{(3)*}=\frac{9(1+\rho)}{4N(1+\sqrt\rho)^2},
\label{eq:triangle_CS}
\end{equation}
%\begin{equation}
 %s^{(3)*}=\frac{27(1-\rho^{3/2})^2} {64N^3(1+\sqrt\rho)^6},
% \label{eq:triangle_s}
%\end{equation}
and $U_*=0$. The separate-estimation ratio remains $(1+\rho)/(1+\sqrt\rho)^2$, approaching $1/2$ as $\rho\to1^-$. This limiting improvement refers to the fixed incident-energy convention.

\subsection{Explicit three-port homodyne settings}

Consider a calibrated operating point $(\phi_{1,*},\phi_{2,*},\phi_{3,*})$ satisfying $\Phi_*=\pi\pmod{2\pi}$. For the optimal input representative in
Eq.~\eqref{eq:triangle_input_phases}, suitable local-oscillator phases are
\begin{equation}
\zeta_1^{\mathrm{opt}}=\frac{\pi}{2},\qquad \zeta_2^{\mathrm{opt}}=\phi_{2,*}+\frac{\pi}{2},\qquad \zeta_3^{\mathrm{opt}}=\frac{\pi}{2}-\phi_{1,*} \pmod{\pi}.
\label{eq:triangle_LO_phases}
\end{equation}
The input and local-oscillator phases are set at this operating point and held fixed when taking derivatives with respect to the unknown phases.

For the representatives displayed in Eq.~\eqref{eq:triangle_LO_phases}, the rotated Jacobian is
\begin{equation}
\Theta_{\mathrm{opt}}D_{\mathrm{opt}}=\mathcal D_0
\begin{pmatrix}
 -1&-\rho&-\sqrt\rho\\
 \sqrt\rho&1&\rho\\
 \rho&\sqrt\rho&1
 \end{pmatrix},
 \label{eq:triangle_real_Jacobian}
\end{equation}
where
\begin{equation}
\mathcal D_0=\frac{\sqrt{N/3}(1+\sqrt\rho)}{1-\rho^{3/2}}>0.
\end{equation}
Changing any local-oscillator phase by $\pi$ reverses the sign of the corresponding quadrature outcome and Jacobian row, without changing the Fisher information.

Substituting Eq.~\eqref{eq:triangle_real_Jacobian} into Eq.~\eqref{eq:homodyne_FIM_article} gives
\begin{equation}
 F^{\hmd}=4\mathcal D_0^2
 \begin{pmatrix}
 a_\rho&c_\rho&c_\rho\\
 c_\rho&a_\rho&c_\rho\\
 c_\rho&c_\rho&a_\rho
 \end{pmatrix}
 =Q_*,
 \label{eq:triangle_homodyne_FIM}
\end{equation}
with
\begin{equation}
a_\rho=1+\rho+\rho^2, \qquad c_\rho=\sqrt\rho+\rho+\rho^{3/2}.
\end{equation}
This equality also follows directly from Eq.~\eqref{eq:homodyne_deficit_article}, since the rotated Jacobian is entirely real.

The triangular realization requires only one balanced homodyne detector per output port and a common phase reference stable relative to the input preparation. A preliminary estimation stage may be used to calibrate both the input phases and the local-oscillator phases for the subsequent local estimation stage.

\section{Discussion and conclusions}
We studied the simultaneous local estimation of $n$ internal phases in an $n$-port ring cavity using multimode coherent light. At the optimal operating point, the input energy is equally distributed among all ports, $|\alpha_j|^2=N/n$, and the internal phases $\phi_j$ and input phases $\theta_j$ satisfy $\phi_j+\theta_{j\ominus1}-\theta_j=\pi\pmod{2\pi}$ for every $j$. This condition does not require the internal phases to be equal; their differences are compensated by the relative phases of the input fields. The optimal probe minimizes both SLD bound $\operatorname{Tr}(Q^{-1})$ and sloppiness $(\det Q)^{-1}$, and the Uhlmann curvature vanishes. Under a consistent resource accounting, simultaneous estimation generally provides a constant-factor advantage over separate estimation rather than a better scaling with the number of parameters~\cite{Gorecki2022}. Our result shows the same behavior: the optimized ring-cavity strategy retains both the $n^2$ dependence on the number of phases and the coherent-state $1/N$ scaling with the mean incident photon number. Its advantage is a reduction of the precision prefactor, arising from the repeated interactions generated by the cavity. Finally, a port-resolved homodyne measurement can attain the optimal QFIM, so the bound is not only formal but can be reached with a concrete measurement.

The improvement originates from the cavity feedback represented by $R$, which sums the contributions produced after different numbers of circulations. For the optimal input mode, the phase acquired during propagation compensates the sign introduced by the beam splitters, so that successive contributions to the phase response add with the same phase.
% The cyclic structure provides a Fourier description of both the input field and the phase response, but the two Fourier indices have different meanings. The input mode $\boldsymbol z_m$ specifies the complex amplitudes prepared simultaneously at the $n$ input ports. All ports receive the same intensity, while different values of $m$ correspond to different relative phases between neighboring inputs. By contrast, $k$ in $\boldsymbol v_k$ labels a possible pattern of small changes in the unknown internal phases. The component $k=0$ represents an equal change of all phases, while the components $k\neq0$ describe nonuniform phase variations around the ring. Any local phase perturbation can be decomposed into these independent patterns. For a fixed Fourier input $\boldsymbol z_m$, the QFIM does not couple different phase-variation patterns, and $\lambda_k^{(m)}$ quantifies the information about the pattern labeled by $k$.
\jy{This enhancement benefits all phase-variation modes: the} inequality $\lambda_k^*\geq\lambda_k^{(m)}$ shows that, among all Fourier input modes, the optimal mode provides at least as much information about every phase-variation pattern. Its improved precision therefore does not result from sacrificing the sensitivity to one phase combination in order to improve another. Cyclic averaging extends the resulting cost comparison from the Fourier input modes to an arbitrary coherent input. At the optimal point, the two conjugate components $k$ and $-k$, which represent opposite Fourier phase progressions, have equal response strengths. Since the Uhlmann curvature is determined by the difference between these paired
responses, their equality gives $U_*=0$.

The main practical advantage of the scheme is that it uses coherent light and local homodyne detection. It does not require squeezed states, fixed-photon-number entangled probes, or nonlocal photon-number-resolving measurements~\cite{modak2026}. The optimal configuration can also accommodate a nonuniform internal phase distribution by changing the input phases instead of rebuilding the cavity. Compared with distributed multiphoton protocols, the preparation and readout are simpler, although the resulting precision remains at the coherent-state $1/N$ scaling. The present result is nevertheless restricted to coherent probes, and it does not exclude a further advantage from squeezed or other nonclassical inputs. The model also assumes a lossless cavity with identical beam-splitter reflectivities and performs local estimation around a calibrated operating point\jy{~\cite{Escher2021,Demkowiczlossy1009}.} The input phases and homodyne local-oscillator phases must therefore be phase locked, and the full $n$-parameter measurement uses a stable external phase reference.

{\color{black} Future extensions of this work concern replacing the coherent probe with nonclassical light. Squeezed or entangled multimode inputs can, in principle, improve the precision beyond the coherent-state scaling identified here, and the ring-cavity geometry, with its repeated interactions and Fourier-mode structure, is well suited to such an investigation. The optimal coherent configuration derived here then serves as a rigorous benchmark: any quantum advantage from a nonclassical probe must be measured against the QFIM spectrum in Eq. (38) and the homodyne-attainable bound in Eq. (53). It would also be interesting to relax the assumptions of lossless and identical beam splitters, and to examine whether the phase-matching condition in Eq. (37) survives in the presence of loss or disorder. Finally, the compatibility of the optimal point suggests that the full $n$-parameter estimation can be performed jointly without an incompatibility penalty; whether this favorable structure persists for nonclassical probes or for more general cavity networks remains an open question.}

 {\color{black} {\textbf{Acknowledgments:}} 
KMM and DKM acknowledge financial support from I-Hub Quantum Technology Foundation, IISER Pune for the Chanakya Doctoral Fellowship grant (I-HUB/DF/2022-23/04). AKP and DKM  acknowledge financial support from Institution of Eminence (IoE), Banaras Hindu University, Varanasi, India for the Bridge Grant. DKM and MGAP acknowledge financial support from the Ministry of Education, Government of India for the SPARC grant (No. SPARC/2025-2026/P4343, dated: 20/05/2026).}

\bibliographystyle{unsrt}
\bibliography{qringbib.bib}

\newpage
\appendix
\section{Ring algebra and response derivative}
\label{app:ring_algebra}
Introduce vectors $\bm a,\bm b,\bm c,\bm d$ for the external and intracavity modes. Equations~\eqref{eq:local_bs_b}--\eqref{eq:propagation_matching} give
\begin{equation}
 \bm d=T(\sqrt\tau\,\bm a-\sqrt\rho\,\bm d),
\end{equation}
and therefore
\begin{equation}
 \bm d=\sqrt\tau(I_n+\sqrt\rho T)^{-1}T\bm a.
\end{equation}
Substitution into $\bm b=\sqrt\rho\bm a+\sqrt\tau\bm d$ yields
\begin{align}
 S_n&=\sqrt\rho I_n+\tau(I_n+\sqrt\rho T)^{-1}T\\
 &= (\sqrt\rho I_n+T)(I_n+\sqrt\rho T)^{-1}.
\end{align}

The finite geometric identity
\begin{equation}
 (I_n+X)\sum_{r=0}^{n-1}(-X)^r=I_n-(-X)^n
\end{equation}
with $X=\sqrt\rho T$ and $T^n=\e^{\ii\Phi}I_n$ gives
\begin{equation}
 R=(I_n+\sqrt\rho T)^{-1}=\frac1{A_n}\sum_{r=0}^{n-1}(-\sqrt\rho)^rT^r,
 \label{eq:finite_R_appendix}
\end{equation}
where
\begin{equation}
 A_n=1+(-1)^{n+1}\rho^{n/2}\e^{\ii\Phi}.
\end{equation}
Because $0<\rho<1$, $A_n$ cannot vanish.

For an invertible matrix $X$, differentiating $XX^{-1}=I$ gives
\begin{equation}
\partial_\mu X^{-1}=-X^{-1}(\partial_\mu X)X^{-1}.
\end{equation}
Applying this identity to $R$ and using $\partial_\mu T=\ii E_\mu T$ gives
\begin{equation}
\partial_\mu R=-\ii\sqrt\rho R E_\mu T R.
\end{equation}
Differentiating $S_n=(\sqrt\rho I_n+T)R$ and simplifying with $(\sqrt\rho I_n+T)-\sqrt\rho(I_n+\sqrt\rho T)=\tau T$ yields
\begin{equation}
\partial_\mu S_n=\ii\tau R E_\mu T R.
\end{equation}
For a fixed input $\bm\alpha$,
\begin{align}
D_{\cdot\mu}=(\partial_\mu S_n)\bm\alpha=\ii\tau R E_\mu TR\bm\alpha=\ii\tau(TR\bm\alpha)_\mu R\bm e_\mu,
\end{align}
which gives Eqs.~\eqref{eq:D_factorization_article} and \eqref{eq:D_column_physical}.

\section{Coherent-state quantum geometry}
\label{app:coherent_geometry}
For a single coherent mode,
\begin{equation}
\lvert\beta\rangle=\e^{-|\beta|^2/2}\e^{\beta b^\dagger}\lvert0\rangle.
\end{equation}
Differentiation with respect to a real parameter $\phi_\mu$ gives
\begin{equation}
\lvert\partial_\mu\beta\rangle=\left[b^\dagger\partial_\mu\beta-\frac12\partial_\mu|\beta|^2\right]\lvert\beta\rangle.
\end{equation}
Since
\begin{equation}
\langle\beta\vert\partial_\mu\beta\rangle=(\partial_\mu\beta)\beta^*-\frac12\partial_\mu|\beta|^2,
\end{equation}
the component orthogonal to $\lvert\beta\rangle$ is
\begin{equation}
\left(I-\lvert\beta\rangle\langle\beta\rvert\right)\lvert\partial_\mu\beta\rangle=(\partial_\mu\beta)(b^\dagger-\beta^*)\lvert\beta\rangle.
\end{equation}

Using $b\lvert\beta\rangle=\beta\lvert\beta\rangle$ and $[b,b^\dagger]=1$, one has
\begin{equation}
\langle\partial_\mu\beta|\left(I-\lvert\beta\rangle\langle\beta\rvert\right)|\partial_\nu\beta\rangle=(\partial_\mu\beta)^*\partial_\nu\beta.
\end{equation}

For the multimode product state $\lvert\psi\rangle=\bigotimes_{j=1}^n\lvert\beta_j\rangle$,
\begin{equation}
\lvert\partial_\mu\psi\rangle=\sum_{j=1}^{n}\lvert\partial_\mu\beta_j\rangle\bigotimes_{\ell\neq j}\lvert\beta_\ell\rangle.
\end{equation}
When the quantum geometric matrix is formed, the terms involving derivatives acting on two different modes are exactly cancelled by $\langle\partial_\mu\psi|\psi\rangle \langle\psi|\partial_\nu\psi\rangle$. Therefore only the single-mode contributions remain:
\begin{equation}
J_{\mu\nu}=\sum_{j=1}^{n}\left[\langle\partial_\mu\beta_j|\partial_\nu\beta_j\rangle-\langle\partial_\mu\beta_j|\beta_j\rangle\langle\beta_j|\partial_\nu\beta_j\rangle\right]=\sum_{j=1}^{n}(\partial_\mu\beta_j)^*\partial_\nu\beta_j.
\end{equation}
Since $D_{j\mu}=\partial_\mu\beta_j$, this gives
\begin{equation}
J=D^\dagger D.
\end{equation}

\section{Existence of the uniform representation}
\label{app:existenceG}
Let $G=\diag(\e^{\ii\gamma_1},\ldots,\e^{\ii\gamma_n})$. At the operating point,
\begin{equation}
(G^\dagger T_0G)_{jk}=\e^{\ii(\phi_{0,j}+\gamma_{j\ominus1}-\gamma_j)}\delta_{k,j\ominus1}.
\end{equation}
Thus Eq.~\eqref{eq:gauge_reduction_article} is equivalent to
\begin{equation}
\gamma_j-\gamma_{j\ominus1}=\phi_{0,j}-\frac{\Phi_0}{n}\pmod{2\pi}.
\label{eq:gamma_recursion_appendix}
\end{equation}
The sum of the right-hand side over one cycle vanishes, so the recursion is consistent. Any two solutions differ by a common phase. The matrix $G$ is chosen at the operating point and held fixed while derivatives are taken; hence $\widetilde D=G^\dagger D$ with no derivative of $G$.

\section{Proof of Theorem~\ref{thm:global_article}}
\label{app:Proof_thm}
\subsection{From arbitrary probes to a modal mixture}

% Convexity of $Q\mapsto\Tr(Q^{-1})$ and concavity of $\log\det Q$ imply
% \begin{equation}
%  \CS(\overline Q)\leq\CS(Q(\bm z)), \qquad s(\overline Q)\leq s(Q(\bm z)).
%  \label{eq:average_costs_article}
% \end{equation}

In uniform coordinates, $P$ commutes with $T$ and $R$. From Eq.~\eqref{eq:D_factorization_article}, cyclic relabeling gives
\begin{equation}
 D(P^r\bm z)=P^rD(\bm z)(P^{\mathsf T})^r.
\end{equation}
Therefore, a cyclic relabeling of input ports and estimated phases acts covariantly,
\begin{equation}
Q(P^r\bm z)=P^rQ(\bm z)(P^{\mathsf T})^r.
\label{eq:Q_covariance_article}
\end{equation}

All matrices on this orbit have the same trace-inverse cost and determinant. Define the cyclic comparison matrix
\begin{equation}
\overline Q(\bm z)=\frac1n\sum_{r=0}^{n-1}Q(P^r\bm z).
\label{eq:Q_average_article}
\end{equation}

For positive-definite matrices, $Q\mapsto\Tr(Q^{-1})$ is convex. Hence
\begin{align}
\CS(\overline Q)\leq\frac1n\sum_{r=0}^{n-1}\CS(Q(P^r\bm z))=\CS(Q(\bm z)).
\label{eq:Q_average_Cs}
\end{align}
Concavity of $\log\det Q$ gives
\begin{align}
\log\det\overline Q\geq\frac1n\sum_r\log\det Q(P^r\bm z)=\log\det Q(\bm z),
\end{align}
and therefore 
\begin{equation}
s(\overline Q)\leq s(Q(\bm z)).
\label{eq:Q_average_s}    
\end{equation}
Singular cases follow by continuity and have divergent cost. The average is a comparison matrix, not the QFIM of a new pure probe. Its purpose is to remove the directional preference associated with choosing a distinguished starting port.

\subsection{Cancellation of cross-winding terms}
Because $D$ is linear in the input,
\begin{equation}
D(\bm z)=\sum_m c_mD^{(m)}, \qquad D^{(m)}:=D(\sqrt N\bm z_m).
\end{equation}
The Fourier eigenvalue equation $P^r\bm z_m=\omega^{rm}\bm z_m$ gives
\begin{equation}
D(P^r\bm z)=\sum_{m=0}^{n-1}c_m\omega^{rm}D^{(m)}.
\label{eq:app_shifted_D_modes}
\end{equation}
Consequently,
\begin{equation}
\begin{aligned}
J(P^r\bm z) &=D(P^r\bm z)^\dagger D(P^r\bm z)\\
&=\sum_{m,m'=0}^{n-1}c_m^*c_{m'}\omega^{r(m'-m)}(D^{(m)})^\dagger D^{(m')}.
\end{aligned}
\label{eq:app_F_cross_terms}
\end{equation}
The finite Fourier orthogonality relation is
\begin{equation}
\frac1n\sum_{r=0}^{n-1}\omega^{r(m'-m)}=\delta_{mm'}.
\label{eq:app_root_orthogonality}
\end{equation}
Averaging Eq.~\eqref{eq:app_F_cross_terms} over $r$ removes every term with $m\neq m'$:
\begin{equation}
\overline J(\bm z)=\sum_{m=0}^{n-1}|c_m|^2(D^{(m)})^\dagger D^{(m)}=\sum_{m=0}^{n-1}|c_m|^2J_m.
\label{eq:app_F_mixture}
\end{equation}
Because the weights $|c_m|^2$ are real, taking the real part commutes with this sum, and hence
\begin{equation}
\overline Q(\bm z)=4\Ree\overline J(\bm z)=\sum_{m=0}^{n-1}|c_m|^2Q_m, \qquad Q_m:=Q(\sqrt N\,\bm z_m).
\label{eq:modal_mixture_article}
\end{equation}
The matrix $\overline Q$ is introduced only for comparison: it is the cyclic average of QFIMs, not the QFIM of the original coherent superposition after its relative mode phases have been discarded.

\subsection{Spectrum of QFIM for fixed probes}
Fix the input $\bm\alpha=\sqrt N\bm z_m$ and define
\begin{equation}
g_r=\frac{1}{1+\sqrt\rho\e^{\ii\Upsilon_r}},\qquad f_r=|g_r|^{-2}=1+\rho+2\sqrt\rho\cos\Upsilon_r.
\label{eq:g_f_article}
\end{equation}
Let $\bm v_k$ denote the corresponding Fourier vector in parameter space with $(\bm v_k)_\mu=\frac1{\sqrt n}\omega^{-k(\mu-1)}$. Although $\bm v_k$ and $\bm z_k$ have the same coordinate form, they have different roles: $\bm z_m$ is an optical input pattern, whereas $\bm v_k$ is a spatial Fourier combination of infinitesimal phase variations. All Fourier indices are understood modulo $n$. Hence $\bm v_{-k}=\bm v_{n-k}$ for $k\neq0$, and entrywise complex conjugation gives $\bm v_k^*=\bm v_{-k}$.

Denote the corresponding matrices by $D^{(m)}$. Since $T\bm z_r=\e^{\ii\Upsilon_r}\bm z_r$,
\begin{equation}
R\bm z_r=g_r\bm z_r,\qquad TR\bm\alpha=\sqrt N\e^{\ii\Upsilon_m}g_m\bm z_m.
\end{equation}
The componentwise product of two normalized Fourier vectors is
\begin{equation}
\diag(\bm z_m)\bm v_k=\frac1{\sqrt n}\bm z_{m+k}.
\end{equation}
Substituting into Eq.~\eqref{eq:D_factorization_article} gives
\begin{align}
D^{(m)}\bm v_k&=\ii\tau\sqrt N\e^{\ii\Upsilon_m}g_mR\diag(\bm z_m)\bm v_k\\
&=\ii\sqrt{\frac Nn}\tau\e^{\ii\Upsilon_m}g_mg_{m+k}\bm z_{m+k}\\
&=\ii\sqrt{\frac Nn}(1-\rho) \e^{\ii\Upsilon_m}g_mg_{m+k}\bm z_{m+k}.
\label{eq:Dvk_article}
\end{align}
Thus a phase perturbation with spatial frequency $k$ shifts the optical winding index from $m$ to $m+k$. Since the output modes are orthogonal, the $\bm v_k$ diagonalize $J^{(m)}=(D^{(m)})^\dagger D^{(m)}$. 

Orthogonality of the output winding modes implies
\begin{equation}
\bm v_k^\dagger J^{(m)}\bm v_{k'}=\delta_{kk'}\frac{N\tau^2}{n}|g_m|^2|g_{m+k}|^2.
\end{equation}
Thus
\begin{equation}
J^{(m)}\bm v_k=\Gamma_k^{(m)}\bm v_k, \qquad \Gamma_k^{(m)}=\frac{N(1-\rho)^2}{nf_mf_{m+k}}.
\end{equation}

Entrywise complex conjugation is denoted by a star. Since
\begin{equation}
Q^m=2(J^{(m)}+J^{(m)*}),\qquad U_m=-2\ii(J^{(m)}-J^{(m)*}),
\end{equation}
and $\bm v_k^*=\bm v_{-k}$,
\begin{equation}
J^{(m)*}\bm v_k=\Gamma_{-k}^{(m)}\bm v_k.
\end{equation}
Therefore
\begin{align}
Q_m\bm v_k&=2(\Gamma_k^{(m)}+\Gamma_{-k}^{(m)})\bm v_k,\\
U_m\bm v_k&=2\ii(\Gamma_{-k}^{(m)}-\Gamma_k^{(m)})\bm v_k.
\end{align}

The resulting QFIM eigenvalues are
\begin{equation}
\lambda_k^{(m)}=\frac{2N(1-\rho)^2}{nf_m}\left(\frac1{f_{m+k}}+\frac1{f_{m-k}}\right),
\label{eq:lambda_mode_article}
\end{equation}
with winding indices understood modulo $n$. 

\subsection{Proof of spectral dominance and equality conditions}
\label{app:dominance_article}
To compare all modes and operating points, regard the effective increment as a continuous variable and write $\delta_k=2\pi k/n$,
\begin{equation}
f(\vartheta)=1+\rho+2\sqrt\rho\cos\vartheta,
\end{equation}
\begin{equation}
\lambda_k(\Upsilon)=\frac{2N(1-\rho)^2}{nf(\Upsilon)}
\left[\frac1{f(\Upsilon+\delta_k)}+\frac1{f(\Upsilon-\delta_k)}\right].
 \label{eq:lambda_continuous_article}
\end{equation}

For a physical mode, $\lambda_k^{(m)}(\Phi)=\lambda_k(\Upsilon_m)$. The central spectral inequality is
\begin{equation}
\lambda_k(\pi)\geq\lambda_k(\Upsilon)\qquad\text{for every }k\text{ and }\Upsilon.
\label{eq:spectral_dominance_article}
\end{equation}
The factor $1/f(\Upsilon)$ is the cavity response of the excited winding mode, while the shifted factors describe the output modes generated by the conjugate parameter directions $k$ and $-k$. The inequality states that their combined contribution is maximized at the same $\pi$ phase matching for every parameter direction. Next we prove Eq.~\eqref{eq:spectral_dominance_article}. Set
\begin{equation}
\eta=\frac{2\sqrt\rho}{1+\rho},\qquad 0<\eta<1,
\end{equation}
and write $f(\vartheta)=(1+\rho)[1+\eta\cos\vartheta]$. Removing the positive factor $2N(1-\rho)^2/[n(1+\rho)^2]$ from Eq.~\eqref{eq:lambda_continuous_article}, define
\begin{align}
\mathcal H_k(\Upsilon)=\frac{1}{1+\eta\cos\Upsilon}\left[\frac1{1+\eta\cos(\Upsilon+\delta_k)}+\frac1{1+\eta\cos(\Upsilon-\delta_k)}\right].
\end{align}
Let
\begin{align}
\Pi_k(\Upsilon)&=[1+\eta\cos(\Upsilon+\delta_k)][1+\eta\cos(\Upsilon-\delta_k)]\\
&=1-\eta^2+\eta^2(\cos^2\delta_k+\cos^2\Upsilon)+2\eta\cos\Upsilon\cos\delta_k.
\end{align}
Then
\begin{equation}
\mathcal H_k(\Upsilon) =\frac{2[1+\eta\cos\Upsilon\cos\delta_k]}{(1+\eta\cos\Upsilon)\Pi_k(\Upsilon)},
\end{equation}
whereas
\begin{equation}
\mathcal H_k(\pi)=\frac{2}{(1-\eta)(1-\eta\cos\delta_k)}.
\end{equation}
Direct subtraction gives
\begin{align}
\mathcal H_k(\pi)-\mathcal H_k(\Upsilon)&=\frac{2\eta(1+\cos\Upsilon)R_\eta(\Upsilon,\delta_k)}{(1-\eta)(1-\eta\cos\delta_k)}\times\frac{1}{(1+\eta\cos\Upsilon)\Pi_k(\Upsilon)},
\label{eq:H_difference_appendix}
\end{align}
where
\begin{align}
R_\eta(\Upsilon,\delta)=&\eta\cos^2\delta+(1-\eta+2\eta\cos\Upsilon)\cos\delta+\eta^2\cos^2\Upsilon +\eta(1-\eta)\cos\Upsilon+1-\eta.
\label{eq:R_eta_appendix}
\end{align}
Every factor in the denominator of Eq.~\eqref{eq:H_difference_appendix} is positive, and $1+\cos\Upsilon\geq0$. It remains to prove $R_\eta\geq0$.

We define $x=\cos\Upsilon$ and $c=\cos\delta$,
\begin{equation}
R_\eta(x,c)=\eta c^2+(1-\eta+2\eta x)c+\eta^2x^2+\eta(1-\eta)x+1-\eta.
\end{equation}

To prove $R_\eta(x,c)\geq0$ on $[-1,1]^2$, fix $x=\cos\Upsilon$ and minimize $R_\eta$ with respect to $c=\cos\delta$. Since
\begin{equation}
\frac{\partial R_\eta}{\partial c}=2\eta c+1-\eta+2\eta x, \qquad \frac{\partial^2R_\eta}{\partial c^2}=2\eta>0,
\end{equation}
$R_\eta$ is a convex quadratic function of $c$. Its unique unconstrained minimum occurs at
\begin{equation}
c_0=-\frac{1-\eta+2\eta x}{2\eta}.
\end{equation}
Moreover,
\begin{equation}
c_0-1=-\frac{1+\eta+2\eta x}{2\eta}\leq-\frac{1-\eta}{2\eta}<0,
\end{equation}
where we used $x\geq-1$ and $0<\eta<1$. Hence $c_0<1$.

Therefore, the minimum over $c\in[-1,1]$ is attained either at $c=-1$ or at $c=c_0$:
\begin{equation}
\min_{-1\leq c\leq1}R_\eta(x,c)=
 \begin{cases}
  R_\eta(x,-1),&c_0\leq-1,\\
  R_\eta(x,c_0),&-1\leq c_0\leq1.
 \end{cases}
\end{equation}

In the first case,
\begin{equation}
R_\eta(x,-1)=\eta(1-x)(1-\eta x)\geq0.
\end{equation}
In the second case, $c_0\geq-1$ implies
\begin{equation}
-1\leq x\leq\frac{3\eta-1}{2\eta}.
\end{equation}
This interval is nonempty only when $\eta\geq1/5$. At the minimum,
\begin{equation}
R_\eta(x,c_0)=\frac{1-\eta}{4\eta} \left[5\eta-1-4\eta^2x^2-4\eta(1-\eta)x\right].
\end{equation}
The expression in square brackets is concave in $x$, so its minimum on the allowed interval occurs at an endpoint. At the two endpoints
it equals
\begin{align}
(1-\eta)(8\eta-1)&>0,\quad\text{for } x=-1,\\
3\eta(1-\eta)&>0,\quad\text{for }x=\frac{3\eta-1}{2\eta}.
\end{align}
Hence $R_\eta(x,c)\geq0$ for all $x,c\in[-1,1]$, which proves Eq.~\eqref{eq:spectral_dominance_article}.

The preceding argument also determines the equality conditions. In the case where the quadratic minimum occurs at its vertex inside $[-1,1]$, we have shown that $R_\eta(x,c)>0$. In the boundary case,
\begin{equation}
R_\eta(x,c)\geq R_\eta(x,-1)=\eta(1-x)(1-\eta x).
\end{equation}
Since $0<\eta<1$, the right-hand side vanishes only at $x=1$. At this value, the quadratic is strictly increasing in $c\in[-1,1]$, so equality additionally requires $c=-1$. Therefore,
\begin{equation}
R_\eta(x,c)=0 \quad\Longleftrightarrow\quad(x,c)=(1,-1).
\end{equation}
The factorization of $H_c(-1)-H_c(x)$ contains the additional factor $1+x$, while its denominator is strictly positive. Consequently,
\begin{equation}
\lambda_k(\pi)=\lambda_k(\Upsilon) \quad\Longleftrightarrow\quad
 \begin{cases}
 \Upsilon=\pi\pmod{2\pi}, & \text{for any }k,\\
 \Upsilon=0\pmod{2\pi}, & \text{if }n\text{ is even and }k=n/2.
 \end{cases}
 \label{eq:spectral_equality_appendix}
\end{equation}
The second case gives equality in only one parameter direction. In particular, for $k=0$, one has $c=\cos\delta_0=1$, and hence
\begin{equation}
\lambda_0(\pi)>\lambda_0(\Upsilon)\qquad \text{whenever }\Upsilon\not\equiv\pi\pmod{2\pi}.
\label{eq:strict_k0_appendix}
\end{equation}
Thus every nonoptimal winding mode satisfies $Q_*-Q_m\succeq0$ with $Q_*-Q_m\neq0$. This strict inequality in the $k=0$ direction is the ingredient needed to establish uniqueness of the optimal input.

\subsection{Completion of the global optimization}
Let $Q_*$ be the winding-mode QFIM at $\Upsilon=\pi$. All $Q_m$ share the orthonormal parameter-space eigenvectors $\bm v_k$, so Eq.~\eqref{eq:spectral_dominance_article} implies
\begin{equation} 
Q_*-Q_m =\sum_{k=0}^{n-1} [\lambda_k(\pi)-\lambda_k^{(m)}(\Phi)]\bm v_k\bm v_k^\dagger \succeq0.
\label{eq:mode_Loewner_article}
\end{equation}
Combining this with Eq.~\eqref{eq:modal_mixture_article} gives
\begin{equation}
Q_*-\overline Q=\sum_m|c_m|^2(Q_*-Q_m)\succeq0.
\label{eq:average_Loewner_article}
\end{equation}
For positive-definite matrices, inversion reverses the Loewner order and the determinant is monotone under it. Together with Eqs.~\eqref{eq:Q_average_Cs} and~\eqref{eq:Q_average_s},
\begin{equation}
\CS(Q_*)\leq\CS(\overline Q)\leq\CS(Q(\bm z)),
\label{eq:CS_global_chain_article}
\end{equation}
\begin{equation}
s(Q_*)\leq s(\overline Q)\leq s(Q(\bm z)).
\label{eq:s_global_chain_article}
\end{equation}
This comparison does not require the generally stronger statement $Q_*\succeq Q(\bm z)$ for the unaveraged input. The equality condition gives uniqueness. Whenever $\Upsilon_m\neq\pi$, the $k=0$ inequality is strict. Any nonzero weight on such a mode makes $Q_*-\overline Q$ nonzero and makes the first inequality in each global chain strict. Equality is therefore possible only for
\begin{equation}
\bm z=\e^{\ii\chi}\bm z_{m_*}, \qquad \Upsilon_{m_*}=\pi,
\end{equation}
where $\chi$ is a common phase.

Summing the phase-matching conditions around the ring gives
\begin{equation}
\Phi_*=n\pi\pmod{2\pi}.
\label{eq:optimal_total_phase_article}
\end{equation}
For $\Phi_*\in[0,2\pi)$, one may choose
\begin{equation}
\Phi_*=\begin{cases}\pi,&n\text{ odd},\\0,&n\text{ even},\end{cases} \qquad m_*=\begin{cases}(n-1)/2,&n\text{ odd},\\n/2,&n\text{ even}.
\end{cases}
\label{eq:optimal_branch_article}
\end{equation}

Returning to the original physical coordinates, the unique optimal input, up to an arbitrary common phase $\chi$, is
\begin{equation}
\bm\alpha_{\mathrm{opt}}=\sqrt N\,\e^{\ii\chi}G\bm z_{m_*}.
\end{equation}
Consequently, the physical input amplitude at port $j$ is
\begin{equation}
\alpha_{\mathrm{opt},j}= \sqrt{\frac{N}{n}} \exp\left[\ii\chi+\ii\gamma_j -\frac{2\pi\ii m_*(j-1)}{n}\right].
\label{eq:physical_optimal_input}
\end{equation}

Writing $\alpha_{\mathrm{opt},j}= |\alpha_{\mathrm{opt},j}|\e^{\ii\theta_j}$, Eq.~\eqref{eq:physical_optimal_input} gives
\begin{equation}
\theta_j=\chi+\gamma_j-\frac{2\pi m_*(j-1)}{n}\pmod{2\pi}.
\label{eq:optimal_input_phase}
\end{equation}
Therefore,
\begin{equation}
\theta_{j\ominus1}-\theta_j=\gamma_{j\ominus1}-\gamma_j+\frac{2\pi m_*}{n}\pmod{2\pi}.
\label{eq:optimal_phase_difference}
\end{equation}
The phase-balancing transformation $G$ satisfies
\begin{equation}
\phi_j+\gamma_{j\ominus1}-\gamma_j= \frac{\Phi}{n}\pmod{2\pi}.
\label{eq:phase_balancing_relation}
\end{equation}
Combining Eqs.~\eqref{eq:optimal_phase_difference} and \eqref{eq:phase_balancing_relation}, one obtains
\begin{equation}
\phi_j+\theta_{j\ominus1}-\theta_j= \phi_j+\gamma_{j\ominus1}-\gamma_j+\frac{2\pi m_*}{n}=\frac{\Phi}{n}+\frac{2\pi m_*}{n}= \Upsilon_{m_*} =\pi\pmod{2\pi}.
\label{eq:optimal_effective_phase}
\end{equation}
Hence the optimal physical input satisfies Eq.~\eqref{eq:physical_optimum_article}.

\section{Derivation of the optimal homodyne response}
\label{app:homodyne_response}

We derive the phase structure of the optimal response matrix and then transform the result back to the original physical coordinates. The calculation is first performed in the uniform cyclic coordinates, where
\begin{equation}
T=\e^{\ii\Phi/n}P.
\end{equation}

At the optimum, the winding index $m_*$ is chosen such that
\begin{equation}
\Upsilon_{m_*}=\frac{\Phi_*+2\pi m_*}{n}=\pi\pmod{2\pi}.
\end{equation}
Consequently,
\begin{equation}
T\bm z_{m_*}=-\bm z_{m_*}, \qquad R\bm z_{m_*}=\frac{1}{1-\sqrt\rho}\bm z_{m_*}.
\end{equation}

The common phase of the optimal input is irrelevant for the QFIM. We therefore first choose the representative
\begin{equation}
\bm\alpha_{\mathrm{uni}}^{\mathrm{opt}}=\sqrt N\,\bm z_{m_*}.
\end{equation}
Using
\begin{equation}
TR\bm\alpha_{\mathrm{uni}}^{\mathrm{opt}} =-\frac{\sqrt N}{1-\sqrt\rho}\bm z_{m_*}
\end{equation}
and $\tau=1-\rho$, Eq.~\eqref{eq:D_factorization_article} gives
\begin{equation}
D_{\mathrm{uni}}^{\mathrm{opt}}=-\ii\sqrt N(1+\sqrt\rho)\,R\diag(\bm z_{m_*}).
\label{eq:D_opt_matrix_appendix}
\end{equation}

At the optimal total phase,
\begin{equation}
\e^{\ii\Phi_*}=(-1)^n.
\end{equation}
Therefore, the finite expansion of the inverse propagator becomes
\begin{equation}
R=\frac{1}{1-\rho^{n/2}}\sum_{r=0}^{n-1}(-\sqrt\rho)^r\e^{\ii r\Phi_*/n}P^r.
\label{eq:R_optimal_appendix}
\end{equation}
For
\begin{equation}
\ell(j,\mu)=(j-\mu)\bmod n,\qquad \ell(j,\mu)\in\{0,\ldots,n-1\},
\end{equation}
the matrix element of $P^r$ satisfies
\begin{equation}
(P^r)_{j\mu}=\delta_{r,\ell(j,\mu)}.
\end{equation}
Hence
\begin{equation}
R_{j\mu}=\frac{(-\sqrt\rho)^{\ell(j,\mu)}\e^{\ii\ell(j,\mu)\Phi_*/n}}{1-\rho^{n/2}}.
\label{eq:R_element_opt_appendix}
\end{equation}

The optimal-mode condition
\begin{equation}
\e^{\ii\Phi_*/n}\omega^{m_*}=-1
\end{equation}
also gives the components of the winding mode in the form
\begin{equation}
(\bm z_{m_*})_\mu= \frac{1}{\sqrt n}(-1)^{\mu-1} \e^{\ii(\mu-1)\Phi_*/n}.
\label{eq:z_opt_components_appendix}
\end{equation}
For fixed $j$ and $\mu$, the definition of $\ell(j,\mu)$ implies
\begin{equation}
\ell(j,\mu)+\mu-1=
    \begin{cases}
        j-1, & j\geq\mu,\\
        j-1+n, & j<\mu.
    \end{cases}
\end{equation}
In the second case, the additional full-cycle factor is
\begin{equation}
(-1)^n\e^{\ii\Phi_*}=1.
\end{equation}
Combining Eqs.~\eqref{eq:R_element_opt_appendix} and
\eqref{eq:z_opt_components_appendix}, one obtains
\begin{equation}
R_{j\mu}(\bm z_{m_*})_\mu=(-1)^{j-1}\e^{\ii(j-1)\Phi_*/n} \frac{(\sqrt\rho)^{\ell(j,\mu)}}{\sqrt n(1-\rho^{n/2})}.
\label{eq:Rz_opt_appendix}
\end{equation}
Substitution into Eq.~\eqref{eq:D_opt_matrix_appendix} yields
\begin{equation}
\left(D_{\mathrm{uni}}^{\mathrm{opt}}\right)_{j\mu}=-\ii(-1)^{j-1} \e^{\ii(j-1)\Phi_*/n} \frac{\sqrt N(1+\sqrt\rho)}{\sqrt n(1-\rho^{n/2})} (\sqrt\rho)^{\ell(j,\mu)}.
\label{eq:D_opt_uniform_appendix}
\end{equation}

For a fixed output port $j$, the factor depending on $\mu$ in
Eq.~\eqref{eq:D_opt_uniform_appendix} is real. Thus all derivatives in the same row have a common phase modulo $\pi$,
\begin{equation}
\widetilde\zeta_j^{\mathrm{opt}}=-\frac{\pi}{2}+\frac{(j-1)\Phi_*}{n}\pmod{\pi}.
\label{eq:LO_uniform_appendix}
\end{equation}
Indeed,
\begin{equation}
\e^{-\ii\widetilde\zeta_j^{\mathrm{opt}}}\left(D_{\mathrm{uni}}^{\mathrm{opt}}\right)_{j\mu}=(-1)^{j-1}\frac{\sqrt N(1+\sqrt\rho)}{\sqrt n(1-\rho^{n/2})}
(\sqrt\rho)^{\ell(j,\mu)}\in\mathbb R.
\label{eq:rotated_D_uniform_appendix}
\end{equation}
This proves that one quadrature per output port is sufficient in the uniform cyclic coordinates.

It remains to express the result in the original physical coordinates. Let
\begin{equation}
G=\diag\left(\e^{\ii\gamma_1},\ldots,\e^{\ii\gamma_n} \right)
\end{equation}
be the fixed diagonal rephasing used to define the uniform cyclic representation. If the arbitrary common phase of the input is denoted by $\chi$, then
\begin{equation}
\bm\alpha_{\mathrm{phys}}^{\mathrm{opt}}=\sqrt N\,\e^{\ii\chi}G\bm z_{m_*}.
\label{eq:physical_opt_input_homodyne_appendix}
\end{equation}
Because $G$ is fixed while the local derivatives are taken, the response matrices in the two coordinate systems are related by
\begin{equation}
D_{\mathrm{phys}}^{\mathrm{opt}}=\e^{\ii\chi}GD_{\mathrm{uni}}^{\mathrm{opt}}.
\label{eq:D_physical_uniform_relation_appendix}
\end{equation}
Therefore, the physical response row has the common phase
\begin{equation}
\zeta_{j,\mathrm{phys}}^{\mathrm{opt}}=\chi+\gamma_j-\frac{\pi}{2}+\frac{(j-1)\Phi_*}{n}\pmod{\pi}.
\label{eq:LO_physical_from_G_appendix}
\end{equation}
The phase $\gamma_j$ appears here because the transformation from uniform to physical coordinates rephases the $j$th output mode. It is not an additional optical element that must be inserted into the experiment; its effect is included by calibrating the physical local-oscillator phase.

To express Eq.~\eqref{eq:LO_physical_from_G_appendix} in terms of the actually prepared input phase, write
\begin{equation}
\alpha_{\mathrm{phys},j}^{\mathrm{opt}}=\sqrt{\frac{N}{n}}\e^{\ii\theta_j}.
\end{equation}
From Eq.~\eqref{eq:physical_opt_input_homodyne_appendix},
\begin{equation}
\theta_j=\chi+\gamma_j -\frac{2\pi m_*(j-1)}{n}\pmod{2\pi}.
\label{eq:theta_opt_homodyne_appendix}
\end{equation}
Subtracting this expression from Eq.~\eqref{eq:LO_physical_from_G_appendix} gives
\begin{equation}
\begin{aligned}
\zeta_{j,\mathrm{phys}}^{\mathrm{opt}}-\theta_j &=-\frac{\pi}{2}+(j-1)\left(\frac{\Phi_*}{n} +\frac{2\pi m_*}{n}\right)\\
&=-\frac{\pi}{2}+(j-1)\Upsilon_{m_*}\\
&=-\frac{\pi}{2}+(j-1)\pi=\frac{\pi}{2}\pmod{\pi}.
\end{aligned}
\end{equation}
Consequently,
\begin{equation}
\zeta_{j,\mathrm{phys}}^{\mathrm{opt}}=\theta_j+\frac{\pi}{2}\pmod{\pi}.
\label{eq:LO_physical_derived_appendix}
\end{equation}
Thus, in the physical implementation, the local oscillator at port $j$ is set to the quadrature phase shifted by $\pi/2$ relative to the prepared input phase at that port. Eqs.~\eqref{eq:D_opt_uniform_appendix}--\eqref{eq:LO_physical_derived_appendix} establish the row-phase condition used in Sec.~\ref{sec:homodyne_attainment}.

\end{document}